\documentstyle[12pt,epsfig]{article}
\begin{document}

\newcommand{\be}{\begin{equation}}
\newcommand{\ee}{\end{equation}}
\newcommand{\ba}{\begin{eqnarray}}
\newcommand{\ea}{\end{eqnarray}}
\newcommand{\tr}{\mbox{tr}}
\newcommand{\GeV}{ \mbox{GeV}}
\newcommand{\TeV}{ \mbox{TeV}}
\newcommand{\rauthors}[1]{#1}
\newcommand{\rjournal}[1]{{\it #1}}
\newcommand{\rvolume}[1]{\mbox{\bf #1}}
\newcommand{\ryear}[1]{\mbox{(#1)}}
\newcommand{\rpage}[1]{\mbox{#1}}
\newcommand{\journref}[5]{\rauthors{#1}\ \rjournal{#2}\ \rvolume{#3}\ %
\ryear{#4}\ \rpage{#5}}
\newcommand{\ch}{\mbox{ch}}
\newcommand{\sh}{\mbox{sh}}
\newcommand{\appex}{ \setcounter{equation}{0}\section}
\newcommand{\figitem}[1]{\item[{\bf Fig.~\ref{#1}}]}
\newcommand{\figcaption}{\refstepcounter{figure}\par\noindent
\centerline{\large\bf Figure \arabic{figure}}\par}

\begin{center}
{\Large\bf Angular and momentum asymmetry in particle production
at high energies.}
\end{center}
\medskip
\begin{center}
{\bf Andrei Leonidov$^{(1,2)}$\footnote{e-mail: leonidov@td.lpi.ac.ru} and
Dmitry Ostrovsky$^{(2)}$\footnote{e-mail: ostrov@td.lpi.ac.ru}}
\end{center}
\medskip
\begin{center}
{\it (1) School for Physics and Astronomy, University of Minnesota,
Minneapolis MN 5455, USA}
\end{center}
\medskip
\begin{center}
{\it (2) P.N.~Lebedev Physical Institute, 117924 Leninsky pr. 53, Moscow,
Russia}
\end{center}
\bigskip
\begin{center}
{\bf Abstract}\\
Angular asymmetry and momentum disbalance for a pair of particles
produced at high energy in central rapidity region are studied.
The asymmetry is substantial for small momenta of produced
particles but diminishes when they rise.
\end{center}
\medskip

\newpage

\section{Introduction}

The QCD description structure of the final state in high energy processes
is one of the most important subjects in strong interaction physics.
Of special interest is the analysis of the final state formation at high
energies, where one should resum the contributions logarithmic in energy
(Bjorken $x$). This problem has drawn a considerable interest and was studied in
a number of recent publications \cite{Marchesini}, \cite{Salam}.

In the approach traditionally applied for describing the final state in the high
energy hadron collisions, that of collinear factorization,
the final state is produced with the zero total transverse momentum.
In particular for two-particle final state this means that these particles
leave the collision point in the opposite directions and having equal
absolute values of their transverse momenta. We will refer to such a
configuration as symmetric and will consider any deviation  from it as
asymmetry. The value of the asymmetry can be used as the measure of the
departure from the collinear factorization showing
limit of its applicability. To quantify the effects related to the nonvanishing
total transverse momentum of the final state one has to use an approach in which
the transverse momenta of incoming partons are not neglected. Let us note, that
in a number of recent experiments
 on diphoton \cite{diphoton}, $\pi^0$ and
direct photon \cite{pi0} production a substantial discrepancy
between the data and  predictions of collinearly factorized NLO QCD
was observed. Taking into account the  intrinsic transverse momenta
of order of $1-2$ GeV substantially improves this  situation. The results
of \cite{pi0} also show sufficiently broad angular distribution in
$\pi^0$ production.

The paper is organized as follows. In the next section the high energy
factorization will be reviewed and compared with the collinear factorization
one. In section \ref{asymmetry} numerical results for the shape of the final
state containing two particles in the central rapidity region are presented.
The derivation of the expressions for the  cross sections used in our
calculations   can be found in the Appendices.
In section \ref{conclusion} we summarize the results and present our
conclusions.

\section{Collinear and high energy factorization}\label{factorization}

Collinear factorization \cite{CollFact} is a method of describing the strong
interaction processes by factorizing the contribution to physical cross sections
into the product of partons distributions $f_a(x,k^2)$ parametrizing both the
nonperturbative information about the hadron and perturbative evolution
starting from some specific initial condition
and perturbative cross sections corresponding to
the scattering of the parton fluxes.
 In this approach the prehistory of colliding partons is entirely determined by
structure functions while partons themselves are supposed
to be on-shell particles.
The cross section for two-particle (jet) production to the lowest
perturbative (Born) order is
\be\label{CFLO}
\frac{d\sigma}{dk^2dy_1dy_2}=
x_1f_a(x_1,k^2)\frac{d\hat{\sigma}_{ab}}{dt}x_2f_b(x_2,k^2),
\ee
where the sum over parton types $a,b$ is assumed and
$x_{1,2}=k_\perp (e^{\pm y_1}+e^{\pm y_2})/\sqrt{S}$, where $\sqrt{S}$ is an
invariant collision energy.

In this approach it is assumed that
\be\label{CFkinematics}
\Lambda_{QCD}\ll k_\perp \sim x\sqrt{S}\sim \sqrt{S}.
\ee
indicating that one can apply perturbative QCD and
that there is only one big logarithm to take into account, that is
$\ln(k^2/\Lambda^2_{QCD})$.
Resummation of the powers of this logarithm
(i.e. $\sim\alpha_s^n\ln^n(k^2/\Lambda^2_{QCD})$) leads to structure
functions depending on $k^2$ and is performed by DGLAP evolution equation
\cite{DGLAP}.

The situation changes at large $S$ when one reaches the kinematic region
\be\label{HEkinematics}
\Lambda_{QCD}\ll k_\perp \sim x\sqrt{S}\ll \sqrt{S}
\ee
Under these conditions another big logarithm
$\ln(1/x)$ exists.  Resummation of powers of this logarithm can
become more important than the one of $\ln(k^2/\Lambda^2_{QCD})$.
The resummation of the leading energy logarithms for the structure
function is described by BFKL equation \cite{BFKL}.

In this kinematical region the transverse momenta of the incoming parton fluxes
can no longer be neglected. To take them into account
a new approach called "$k_\perp$ factorization" was proposed
in \cite{CFM}. Extensive description of the method
and  various applications can be found in \cite{CCH}.
Let us note  that this method was {\it de facto} used earlier in \cite{LR}.

The method of $k_\perp$ factorization is based on considering
"partons" with nonzero transverse momentum and being, in contrast to the
collinear factorization case, off-mass shell particles. Calculation
of physical cross-sections requires a generalization of the
conventional picture of parton flux as described by usual parton
structure functions. To this aim an unintegrated gluon distribution
$\phi(x,q_\perp, k)$ is introduced:
\be\label{phi}
xg(x,k^2)=\int\limits^{k^2}\frac{dq^2_\perp}{q^2_\perp} \phi(x,q_\perp, k),
\ee
where $\phi/q^2$ is proportional to the probability of finding the incident
parton with longitudinal momentum component $xp_a$ ($p_a$ is a momentum
of initial particle) and transverse momentum $q_\perp $. It is
important to stress that unintegrated structure function depends not only
upon the momentum of particle to which it corresponds, $q$, but also on the
global off-mass-shellness of the process, $k^2$.
This is in fact a general property of distributions in quantum theory --
the probability of a particular event depends not only of its
parameters, but also on the global characteristics of the event ensemble.

The particular interrelation between the unintegrated and
integrated structure functions depends on the evolution
equation that the integrated distribution solves.
For the DGLAP evolution \cite{DGLAP} the unintegrated structure function
can be expressed through the integrated one as follows:
\cite{KMR}
\be\label{phiT}
\phi(x,q_\perp,k)=\left(\frac{\alpha_s}{2\pi}\int\limits_x^{1-q_\perp/k}
\frac{dz}{z} P_{gg}(z) xg\left(\frac{x}{z},q_\perp^2\right)\right)
T_g(q_\perp,k)
\ee
where $T_g$ is the Sudakov form factor \cite{MW} for gluon
\be\label{Tfactor}
T_g(q_\perp,k)=\mbox{exp}\left(-\int\limits_{q_\perp^2}^{k^2}
\frac{\alpha_s(p_\perp)}{2\pi}\frac{dp^2_\perp}{p^2_\perp}
\sum\limits_{i=g,q,\bar{q}}\int\limits_0^{1-q_\perp/k} P_{ig}(z) dz\right)
\ee
(for $P_{gg}$ in the last expression one should substitute
$P_{gg}(z)\rightarrow zP_{gg}(z)$ in account of gluons identity).
In a double logarithmic  approximation Eqs.~(\ref{phiT}-\ref{Tfactor})
coincide with the DDT formula \cite{DDT}.

For BFKL evolution \cite{BFKL} the correspondence between integrated
and unintegrated structure functions takes a simpler form
\be\label{phi_BFKL}
\phi(x,q_\perp)\equiv\phi(x,q_\perp,q_\perp)=
\frac{\partial xg(x,q^2_\perp)}{\partial \ln q_\perp^2}.
\ee
This results from  the absence of strong $q_\perp$ ordering in
BFKL ladder.

From the practical point of view, Eqs.~(\ref{phiT}-\ref{Tfactor}) differ
substantially from Eq.~(\ref{phi_BFKL}) only for $q_\perp\ll k$
implying the presence of a big logarithmic contribution
$\propto \ln{k^2/q^2_\perp}$. Therefore we can in fact use
Eqs.~(\ref{phiT}-\ref{Tfactor}) for both types of
QCD evolution.

The calculation of a cross section with virtual colliding particles
presents a highly nontrivial problem.
The first difficulty lies in the correct summation over
polarization states of the incident off-shell
"partons". Another problem is the presence of bremsstrahlung
from initial and final states of colliding particles.
The derivation of the relevant cross sections is outlined in Appendix A.

More specifically, the transverse (high energy) factorization
is based on considering the  $2\rightarrow n+2$ process cross section
with a large rapidity gaps between the two final particles providing a full
rapidity interval for the process under consideration which are almost collinear
to the incident ones
and $n$ particles emitted into the central rapidity region
(so called quasi-multy-Regge kinematics or QMRK \cite{FL1,FL}).
Note that only the amplitude with on-shell and physically polarized in- and
outgoing
particles has precise gauge invariant meaning and only such expressions
can be used in calculating the cross section.
Large rapidity gaps provide a separation between quantities related
to the incident particles and those describing the  hard cross section
of parton production in the central region.

Let us for example consider the process $gg\rightarrow ggg$  in the limit of
high energy
\begin{center}
\vspace*{0cm}
\epsfig{file=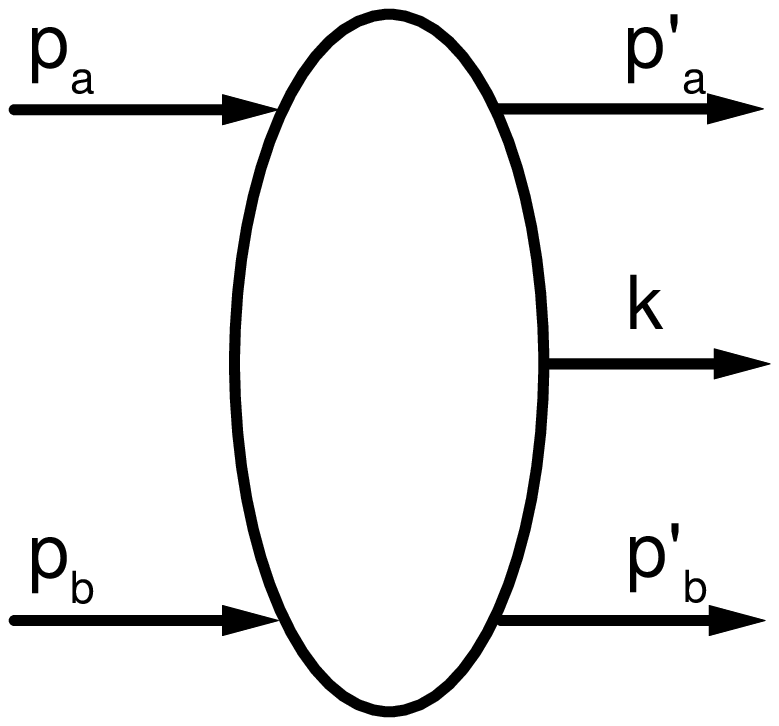,height=35mm}\\
$gg\rightarrow ggg$
\end{center}

\be
\frac{d\sigma_{gg\rightarrow ggg}}{d^2k_\perp dy}=
\frac{4N^3_c\alpha^3_s}{\pi^2 (N^2_c-1)}
\int\frac{d^2q_{1\perp}}{q^2_{1\perp}}
\frac{\delta^{(2)}(q_{1\perp}+q_{2\perp}-k_{\perp})}{k^2_\perp}
\frac{d^2q_{2\perp}}{q^2_{2\perp}}
\ee
with $q_{1,2}=p_{a,b}-p^\prime_{a,b}$.

This equation demonstrates the above mentioned factorization of the cross
section.
The first, second and third factors under the integral  correspond to
$p_a\rightarrow p^\prime_a,q_1$ splitting, $q_1, q_2\rightarrow k$
"scattering" and $p_b\rightarrow p^\prime_2,q_2$ splitting respectively.

The factors related to the splitting of the incident particles
should further be converted to structure functions. One can do it in two steps.
The first step, transformation to the
form factors, is quite straightforward \cite{BG}.
In the second step we have to account for corrections due to radiation along the
directions of incident particles and
replace form factors by unintegrated structure functions $\varphi(x,q_\perp,k)$
with $x$ determined by kinematics.

The cross sections for producing $n=0,1,2$ particles in the central region read

\be\label{sigma0}
\sigma_0=\frac{(2\pi)^2}{N_c^2-1}\int d^2q_\perp
\frac{\varphi(x,q_{\perp})}{q^2_{\perp}}
\frac{\varphi(x,q_{\perp})}{q^2_{\perp}}
\qquad x=q^2_{\perp}/S;
\ee
$$
\frac{d\sigma_1}{d^2k_\perp dy}=
          \int d^2q_{1\perp} d^2q_{2\perp}
               \frac{\varphi(x_{1,0},q_{1\perp},k_\perp)}{q^2_{1\perp}}
               \frac{d\hat{\sigma}_1}{dk^2_\perp}
               \frac{\varphi(x_{2,0},q_{2\perp},k_\perp)}{q^2_{2\perp}},
$$
\be\label{sigma1}
\frac{d\hat{\sigma}_1}{dk^2_\perp}=\frac{4N_c\alpha_s}{N_c^2-1}
\frac{\delta^{(2)}(q_{1\perp}+q_{2\perp}-k_{\perp})}{k^2_{\perp}},
\ee
$$
      x_{1,0}=k_\perp e^{y}/\sqrt{S},\quad x_{2,0}=k_\perp e^{-y}/\sqrt{S};
$$
\begin{figure}[h]
\begin{center}
\vspace*{0cm}
\epsfig{file=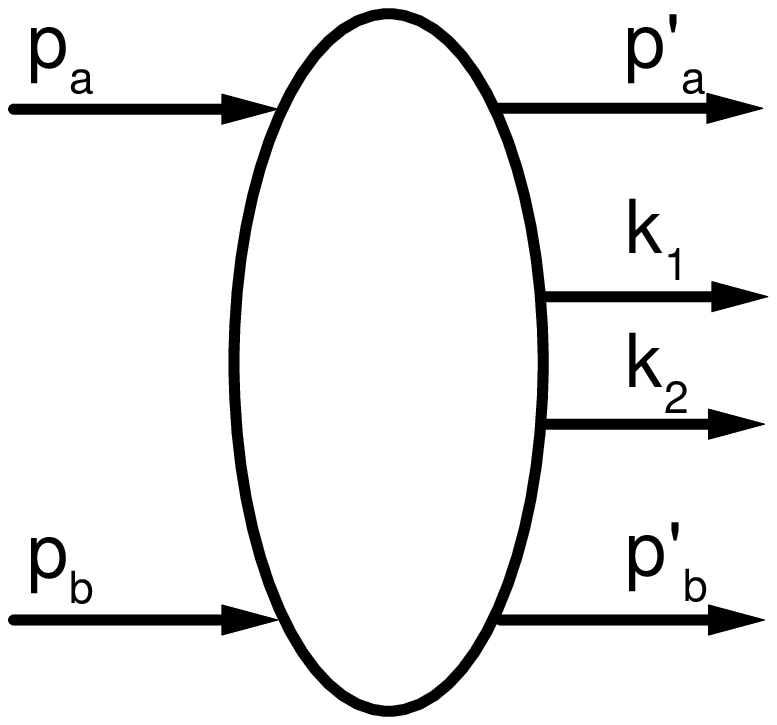,height=35mm}\\
$gg\rightarrow gggg$
\end{center}
\end{figure}
$$
\frac{d\sigma_2}{d^2k_{1\perp} d^2k_{2\perp}dy_1 dy_2}=
        \frac{1}{\pi^2}
        \int d^2q_{1\perp} d^2q_{2\perp}
        \frac{\varphi(x_{1},q_{1\perp},k_{\perp})}{q^2_{1\perp}}
        \frac{d\hat{\sigma}_2}{d^2k_{1\perp} d^2k_{2\perp}}
        \frac{\varphi(x_{2},q_{2\perp},k_{\perp})}{q^2_{2\perp}},
$$
$$
k_\perp=k_{1\perp}+k_{2\perp}
$$
\be\label{sigma2}
\frac{d\hat{\sigma}_2}{d^2k_{1\perp} d^2k_{2\perp}}=
\frac{2N_c^2\alpha_s^2}{N_c^2-1}
\frac{\delta^{(2)}(q_{1\perp}+q_{2\perp}-k_{1\perp}-k_{2\perp})}
{q^2_{1\perp} q^2_{2\perp}} \cal{A},
\ee
$$
x_1=(k_{1\perp}e^{y_1} + k_{2\perp}e^{y_2})/\sqrt{S},\;
x_2=(k_{1\perp}e^{-y_1} + k_{2\perp}e^{-y_2})/\sqrt{S}.
$$
The expression for $n=0$ corresponds to the total cross section in the two gluon
exchange (Low-Nussinov) approximation.
Let us note that the kinematics of Eq.~(\ref{HEkinematics})
is not quite the one in Eq.~(\ref{sigma0}). Having no $k_\perp $
in this process we have to restrict directly $q_\perp $.
In Eq.~(\ref{sigma0}) $q_{\perp}=\sqrt{xS}$
while it is $q_{\perp}\sim x\sqrt{S}$ that is given by
(\ref{HEkinematics}).

The case of one gluon production in the central region ($n=1$) was studied
in a number of publications, e.g. \cite{BG}, \cite{ELR}.

A part of the analytical expression for $\cal{A}$ (for $gg\rightarrow gg$
subprocess) has recently been published in \cite{FL,FKL}.
A similar quantity for $gg\rightarrow q\bar{q}$
subprocess can be found in \cite{FL,FFFK}.  The derivation  of the explicit
expressions is outlined in the Appendices  A and B. Let us note that the formula
for $gg\rightarrow q\bar{q}$ cross section
from \cite{FFFK} coincides
with the analogous formula in \cite{CCH} (in the limit of massless quarks)
obtained by direct application of $k_\perp$ factorization.

For practical applications of the equations Eq.~(\ref{sigma0}-\ref{sigma2}) we
should understand the overall normalization of the off-shell cross sections
to the usual on-shell one arising in collinear factorized formalism.
The normalization can be deduced from
considering the small $q_\perp$ limit in Eq.~(\ref{sigma2}). Physically, we
have to return to Eq.~(\ref{CFkinematics}) and choose the structure functions
being extremely narrow functions of $q_\perp $ and integrate out
$d^2q_{1\perp}$ and $d^2q_{2\perp}$. There is no contradiction with
Eq.~(\ref{CFkinematics}) because $q_\perp\ll k_\perp$.
Certainly, we must put $q_{1\perp},q_{2\perp}=0$
in $\hat\sigma_2$ Eqs.~(\ref{sigma2}, \ref{gggg}, \ref{ggqq}).
As it is follows from Appendix A, $\cal{A}$ is proportional to
$q^2_{1\perp}q^2_{2\perp}$ so this limit for $\hat\sigma_2$ is correct.
After this substitution and integration over $k_{2\perp}$ Eq.~(\ref{CFLO})
appears with the structure functions from Eq.~(\ref{phi}) and with the correct
$d\hat\sigma/dt$.
Consequently, normalization of Eqs.~(\ref{sigma0}-\ref{sigma2})
is also correct.
Let us emphasize that the integration over
$d^2q_{1\perp}$ and $d^2q_{2\perp}$ includes averaging over
angular orientations of $q_{1\perp}$ and $q_{2\perp}$ in the transverse plane
 when arriving to the final expression for  $\hat\sigma_2$.
This averaging is similar to the averaging over initial gluon polarizations
in getting the usual expression for the cross section  of the $2\rightarrow 2$
elastic scattering.

\section{Angular and momentum asymmetry of parton production at
central rapidity}\label{asymmetry}

The most interesting predictions of the high energy factorization are
of course those going beyond the collinear
factorization framework. The simplest situation corresponds
to the production of a single particle
in central rapidity region described in the first order in $\alpha_s$
Eq.~(\ref{sigma1}). This process was first studied in \cite{BG} and
later in many publications, see e.g.  \cite{ELR}.

In this paper we concentrate our attention on another
important feature of the high energy factorization.
From Eq.~(\ref{sigma2}) it is obvious that, in contrast with
the collinear factorization case, the two partons produced in the central
rapidity region are not necessarily going into opposite directions
(back-to-back).
Another important issue  is that the absolute values
of the momenta of outgoing partons are not necessarily equal.
We will refer to these properties as angular and momentum
asymmetries respectively.

The same characteristics of the two-particle production were
studied in \cite{DSS} in the case, where the produced particles are
separated by the large rapidity gap. The angular and momentum asymmetries
considered in \cite{DSS} are due to the presence of BFKL ladder filling the
gap between the produced particles. Let us stress, that in this paper
we study the two-particle decorrelation in a relatively small central
rapidity interval. The rapidity gap, discussed in  \cite{DSS},
is thus absent and decorrelation effects are exclusively due to
off-shellness of the two generalized parton fluxes merging in
a particle-producing vertex.

Our approach is similar to that in \cite{GLR}, where
a multy-regge limit for ${\cal A}$ was considered,
${\cal A}=q^2_{1\perp}q^2_{2\perp}/k^2_{1\perp}k^2_{2\perp}$ (cf. Appendix B).
In this form ${\cal A}$ has no {\it intrinsic} angular and momentum
correlations (see below). Moreover, the contributions from ${\cal A}$  and
the structure functions to $\sigma_2$ (see Eq.~(\ref{sigma2})) factorize
and $\sigma_2$ depends on structure functions only via
gluon-gluon luminosity function \cite{KMR}. The latter contains
practically all information about particle correlations.

A large number of independent variables in Eq.~(\ref{sigma2}) makes it
difficult to visualize the pattern of particles emission.
Therefore we will use the less differential ones, in particular the angular
asymmetry
\be\label{angle}
\left.\frac{d\sigma_2}{d\Delta\phi\,dy_1 dy_2}\right|_{k_0}=
\pi\int\frac{d\sigma_2}{d^2k_{1\perp} d^2k_{2\perp}dy_1 dy_2}
dk_{1\perp }^2 dk_{2\perp }^2,
\ee
where $\Delta\phi\in [0,\pi ]$ is the angle between particles.
Integration over
$k_{1\perp }$ and $k_{2\perp }$ requires introducing the infrared cutoff
limit $k_0$.
The shape of asymmetry is sensitive to the chosen value of $k_0$.
In the normalization of Eq.~(\ref{angle}) and in forthcoming
Eq.~(\ref{momentum}) we took into account
particles identity (even for quarks we do not distinguish between
$q$ and $\bar{q}$).

In order to perform actual calculations one needs to choose the
unintegrated structure functions entering Eqs.~(\ref{sigma0}-\ref{sigma2}).
In principle they should be chosen as solutions to the NLO BFKL equation
\cite{NLOBFKL} or its nonlinear generalizations \cite{nonl}
(also taking to account form factors arising from non-compensating real
and virtual corrections). Leaving this analysis for the future, let us note,
that the minimal requirement surely has to be that whatever expression for
the unintegrated distribution is used, it should not contradict the
information from deep inelastic scattering, i.e. that on integrated
structure function. This fixes the distribution
in the limit $q_\perp=k_\perp$, where  $\phi(x, q_\perp, q_\perp)=\phi(x,
q_\perp)$ (cf. Eq.~(\ref{phi_BFKL})).
In our numerical calculations we have used the unintegrated gluon structure
functions corresponding to integrated CTEQ5M \cite{CTEQ} and AKMS
\cite{AKMS} fitted as in \cite{ELR}.
For $q_\perp<k_\perp$ the unintegrated structure function
$\phi(x, q_\perp, k_\perp)$ was determined according to
Eqs.~(\ref{phiT}),~(\ref{Tfactor}). In the final distributions
we sum over gluon and quark ($n_f=5$) pair contributions
(in fact, the contribution due to quark pair production
is less than $3\%$ of that due to gluons).

In Fig.~\ref{ff3d} we show differential cross section (\ref{angle})
with $y_{1,2}=y_0\mp\Delta y/2$ for $y_0=0$ and $\Delta y\in [0,2]$
and with $k_0=2\mbox{GeV}$.
From now on figures marked by
{\bf a, b} are calculated with CTEQ structure functions,
marked by {\bf c, d} are calculated with AKMS structure functions;
figures with labels {\bf a, c} correspond to $\sqrt{S}=1.8\TeV$, while
those with labels {\bf b, d} correspond to $\sqrt{S}=14\TeV$.

The most striking feature of this cross section that
differs it from the collinear factorized one is the appearance of the
collinear singularity at $\Delta y, \Delta\phi \rightarrow 0$. This
singularity is just a well-known s-chanel one  and
is in turn a reflection of the presence of one particle production
process Eq.~(\ref{sigma1}).
Moreover, the behavior of the two-particle production (\ref{sigma2}) near
the above-described collinear singularity can be presented in the following
decomposed form. Introducing $r^2=\Delta y^2+\Delta\phi^2$,
$k_\perp=k_{1\perp}+k_{2\perp}$, $z=k_{1}/k$ (since $k_1$ and $k_2$ are
collinear this definition is unambiguous) and taking the limit
$r\rightarrow 0$, we get
$$
\frac{d\sigma_2}{d^2k_\perp dy_0 dr dz}=
\frac{d\sigma_1}{d^2k_\perp dy_0}\frac{1}{r}\frac{\alpha_s}{\pi}
(P_{gg}(z)+2n_fP_{qg}(z)),
$$
with one particle production cross section as defined in (\ref{sigma1}).
$P_{gg}$ and $P_{qg}$ are the standard Altarelli-Parisi kernels
\cite{DGLAP} (the factor of 2 before $P_{qg}$
is due to identical treatment of quarks and antiquarks in our
approach).
From Fig.~\ref{ff3d} we can conclude, that the dependence of the asymmetry on
the structure function is quite pronounced,
see corresponding upper and lower panels.
For AKMS structure functions the angular distribution is noticeably
wider then for CTEQ ones. This is the reflection of the fact that AKMS
structure function is broader then CTEQ one if compared at some fixed value of
$x$. The dependence of angular asymmetry on c.m.s. collision energy is,
as seen from Fig.~\ref{ff3d}, relatively weak.

In Fig.~\ref{rhoy} we take a closer look at the angular asymmetry by plotting
the two-dimensional cross sections of the plots in Fig.~\ref{ff3d} for fixed
values of $\Delta y$, i.e. study the
dependence of (\ref{angle}) on $\phi$ for different values of $\Delta y$ where
$$
\rho(\Delta\phi)\sim\frac{d\sigma_2}{d\Delta\phi\,dy_1 dy_2}
$$
is normalized according to
\be\label{rho_norm}
\int\frac{d\Delta\phi}{\pi}\rho(\Delta\phi)=1
\ee
for each value of $\Delta y$. A substantial deviation of $\rho$ from
the back-to-back shape, for which $\rho\sim\delta (\Delta\phi-\pi)$,
is obvious.
From Fig.~\ref{rhoy} we see that the asymmetry grows with increasing
collision energy, although the growth is not pronounced. Using different
structure functions affects the shape of the asymmetry
distribution much stronger than changing the collision energy.

When the cutoff $k_0$ increases, the  kinematic interval
Eq.~(\ref{HEkinematics})
narrows and the two outgoing particles tend to appear back-to-back.
This dependence is clear from Fig.~\ref{rhok} where $\rho(\Delta\phi)$
is shown for $\Delta y=1$ and several values of $k_0$.

We see that when $k_\perp\gg q_{char}$, where $q_{char}$ is a characteristic
transverse momentum carried by the parton fluxes, jets are predominantley
produced at  small $\delta\phi=\pi-\Delta\phi$. This regime  was studied in
DLA in \cite{DDT}. According to \cite{DDT} the (normalized) azimuthal asymmetry
in $\delta\phi\ll 1$ domain is given by
\be\label{rhoDDT}
\rho(\delta\phi)\sim \frac{1}{\delta\phi}\frac{d}{d\delta\phi}T^2_g(\delta\phi)
\ee
with $T_g(\delta\phi)=T_g(\delta\phi\, k_0, k_0)$ (cf. Eq.~(\ref{Tfactor})).
From Eqs.~(\ref{Tfactor}) and (\ref{rhoDDT}) it is obvious that the rapid growth of
$\rho$ with diminishing $\delta\phi$ stops at $\delta\phi^*\approx
exp(-\pi/6\alpha_s)$. Numerically, for $\alpha_s=0.18$ (this value of strong
coupling was used in CTEQ5M calculation for $k=20\GeV$), $\delta\phi^*\approx
5\cdot 10^{-2}$, in agreement with Fig.~\ref{rhok}.
Note that contrary to DLA in our approach the value of $\rho$ for
$\delta\phi<\delta\phi^*$ is not decreasing. The reason is that
in our case there is no scaling leading to Eq.~(\ref{rhoDDT}).
In Fig.~\ref{rhok} the effect of saturation is manifest not for
$\rho(\delta\phi)$ itself, but for its derivative, i.e. for the
corresponding characteristic angle.

Turning now to the analysis of the momentum asymmetry let us consider
the cross section integrated over the angular variables,
\be\label{momentum}
\frac{d\sigma_2}{dk_{1\perp }dk_{2\perp }dy_1 dy_2}=
2k_1k_2\int\frac{d\sigma_2}{d^2k_{1\perp} d^2k_{2\perp}dy_1 dy_2}
d\phi_1 d\phi_2.
\ee
For $y_{1,2}=\pm 0.5$ and $k_{1,2\perp}\in [2,20]$\,GeV this cross section is
plotted in Fig.~\ref{mm}. From this figure we see that,
in a characteristic event,  $k_{1\perp}$ and
$k_{2\perp}$ are not equal. The dominant trend can be described as a
decrease of the cross section with growing $k_{\perp}$.
However, for the values of $k_{1,2\perp}$ of order of, or higher than
$10\,{\rm GeV}$, the correlation pattern is no longer powerlike.
Finally, let us note that the momentum asymmetry shows a significant
dependence on the choice of the structure function and on
c.m.s. collision energy.

\section{Conclusions}\label{conclusion}

Angular and momentum asymmetry is a characteristic feature of
particle production in semihard kinematic region. When
the momenta of produced particles are of the same order as the
characteristic transverse momenta carried by the generalized
off-shell partonic fluxes described by the unintegrated structure
function, the angular distribution of particles shows significant
deviations from the conventional back-to-back picture. The
asymmetry dies away when the transverse momenta of produced particles
are, correspondingly, much larger than the intrinsic ones.
The momentum asymmetry pattern  is somewhat
different due to rapid growth of the cross section with decreasing
momenta.
Nevertheless some reflection of momentum balance in the case of
relatively high momenta is shows itself through local cross
section deviation  from power-like regime
around the point where  the transverse momenta of produced particles
are equal.

\begin{center}
\it Acknowledgements
\end{center}

We are grateful to L. McLerran, Yu. Dokshitzer and Yu. Kovchegov for useful and
stimulating discussions. A.L. is grateful to L. McLerran for kind hospitality at
the University of Minnesota.

We are grateful to O.V.~Ivanov for pointing us out a powerful method of
multidimensional integration \cite{Korobov}.

The work of D.O. was partially supported by the INTAS within the research
program of ICFPM, grant 96-0457.


\setcounter{section}{0}
\renewcommand{\thesection}{{\it Appendix} \Alph{section}}
\renewcommand{\theequation}{\Alph{section}.\arabic{equation}}
\renewcommand{\thesubsection}{\Alph{section}.\arabic{subsection}}

\appex{Particle production in QMRK}

As mentioned before the QMRK regime is that in which the incident particles
scatter at parametrically small angle producing particle(s) in the central
rapidity region.
The leading contribution to the scattering amplitude in this
kinematics has the form (in the Feynman gauge) \cite{FL}
\be\label{decomposition}
A_{2\rightarrow n+2}=g^2\Gamma^{i_1}_a \frac{1}{q^2_{1}}p_a^{\mu_1}
                         M^{i_1 i_2}_{\mu_1 \mu_2}
\renewcommand{\thesubsection}{\Alph{section}.\arabic{subsection}}
                         p_b^{\mu_2}\frac{1}{q^2_{2}}\Gamma^{i_2}_b,
\ee
where $\Gamma$ is a (helicity conserving) vertex and  $i$ stands for the
adjoint representation index. The incident  particles have initial momenta $p_a$
and $p_b$ and the final ones
 $p_a^\prime=p_a-q_1$ and $p_b^\prime=p_b-q_2$:
$$
p_a^\prime=(1-x_{1})p_a-q_{1\perp}-\frac{2q^2_{1\perp}}{(1-x_{1})S}p_b,\quad
p_b^\prime=(1-x_{2})p_a-q_{2\perp}-\frac{2q^2_{2\perp}}{(1-x_{2})S}p_a,
$$
where $p_ap_b=S/2$.
In QMRK approximation one neglects the terms proportional to $p_b$ in
$p_a^\prime$ and proportional to $p_a$ in $p_b^{\prime}$.
Now
$$
q_1=x_{1}p_a+q_{1\perp},\quad q_2=x_{2}p_b+q_{2\perp}
$$
and $q_1^2=q^2_{1\perp}$ and $q_2^2=q^2_{2\perp}$

The explicit expression for $\Gamma$ depends on the nature of incident
particles. For example, if the incident particle $a$ is a gluon the vertex has a
from
$$
\Gamma^{i}_a=2f^{i}_{aa^\prime}g_{\alpha\alpha^\prime}
\epsilon^{\alpha}(p_a) \epsilon^{\alpha^\prime}(p^\prime_a),
$$
where $f^{i}_{aa^\prime}$ is gauge algebra structure constant and
$\epsilon$ is the gluon polarization vector.
For quark scattering
$$
\Gamma^{i}_a=2t^{i}_{AA^\prime}\bar{u}(p_a)\frac{\not{p_b}}{S}v(p^\prime_a),
$$
where $t^{i}_{AA^\prime}$ is now a matrix in fundamental representation.

In general, color structure of (\ref{decomposition}) can be presented as
$T^{i_1}_a T^{i_1 i_2}_{\dots}T^{i_2}_b $ where different $T$'s are
color algebra generators in appropriate representation ($\dots$ denote
color indices of particles produced in the central rapidity region).
The corresponding factor in the cross section reads
$$
\tr (T^{i_1}_a T^{i^\prime_1}_a)
\tr (T^{i_1 i_2}_{\dots} T^{i^\prime_1 i^\prime_2}_{\dots})
\tr (T^{i_2}_b T^{i^\prime_2}_b)
$$
Using the well known property of irreducible representations ,
$\tr (T^{i_1}_a T^{i^\prime_1}_a)\sim\delta^{i_1 i^\prime_1}$,
the summation over final and averaging over initial color indices
can be converted into the averaging over $i_1$ and $i_2$
indices in $M^{i_1 i_2}_{\mu_1 \mu_2}$ in (\ref{decomposition}) with
appropriate  factors included into $\Gamma$'s (and thus into  structure
function definition). As these additional factors  are
completely independent of the structure of $M$ we are having unambiguous
determination of the factorization of the cross section
into structure functions and a generalized cross section for the scattering of
virtual particles described by them. The correct normalization is, in
particular, crucial for getting a correct limit of collinear factorization in
which the hard cross section for the scattering of on-shell particles described
by the usual ("integrated") structure functions should have correct color
factors built in.

When particles produced include gluons the amplitude $M_{\mu_1 \mu_2}$
gets contributions not only from diagrams of $2 \rightarrow n+2$ type
with $n$ lines attached to the $t$-channel gluon,  but also
from diagrams with bremsstrahlung from $p_a$ ($p^\prime_a$)
and $p_b$ ($p^\prime_b$)lines . These can be written in a form of
(\ref{decomposition}) but, having no gluon with momentum $q_1$ ($q_2$)
in the $t$-chanel, give contribution to $M$ proportional to
$q^2_{1\perp}$ ($q^2_{2\perp}$).
Thus the amplitude $M_{\mu_1\mu_2}$ has the form
$$
M_{\mu_1\mu_2}=M^{(1)}_{\mu_1\mu_2}+
\frac{q^2_1}{x_{1}x_{2}S}M^{(2)}_{\mu_1\mu_2}+
\frac{q^2_2}{x_{1}x_{2}S}M^{(3)}_{\mu_1\mu_2}+
\frac{q^2_1 q^2_2}{(x_{1}x_{2}S)^2}M^{(4)}_{\mu_1\mu_2},
$$
Note that if even one of $n$ particles is a gluon produced by bremsstrahlung
from $p_a$ (or $p_a^\prime$) the corresponding diagram
contributes to $M^{(2)}$. In the collinear factorization limit only
$M^{(1)}$ contribution survives.

Our next goal is to show that the amplitude $M$ can be rewritten in such a way,
that the nonsense polarizations dominating the fluxes coming to the hard
vertex can effectively be traded for the transverse ones providing a basis for
interpreting the hard block contribution as a (modified) cross section.
To do this let us  consider the amplitude
$A_{2\rightarrow n+2}$ in the  axial gauge
with the gauge vector lying in $(p_a,p_b)$ plane, $n=ap_a+bp_b$.
In this gauge $\Gamma$'s do not change and the numerator of the gluon propagator
is
$$
d_{\mu\nu}(q)=g_{\mu\nu}-\frac{n_\mu q_\nu + q_\mu n_\nu}{n\cdot q}+
n^2\frac{q_\mu q_\nu}{(n\cdot q)^2}.
$$
It is straightforward to check that the following important relations hold
\begin{equation}\label{projection}
p_a^\mu d_{\mu\nu}(q_1)=-\frac{1}{x_{1}}q_{1\perp, \nu}, \quad
p_b^\mu d_{\mu\nu}(q_2)=-\frac{1}{x_{2}}q_{2\perp, \nu}.
\end{equation}

Let us now  inspect how the structure of
$p_a^{\mu_1}M_{\mu_1\mu_2}p_b^{\mu_2}$ changes in this gauge
\begin{eqnarray}
p_a^{\mu_1}M^{(1)}_{\mu_1\mu_2}p_b^{\mu_2}
&\rightarrow&
p_a^{\mu_1}d_{\mu_1}^{\nu_1}(q_1)M^{(1)}_{\nu_1\nu_2}
d_{\mu_2}^{\nu_2}(q_1)p_b^{\mu_2},\nonumber \\
p_a^{\mu_1}M^{(2)}_{\mu_1\mu_2}p_b^{\mu_2}
&\rightarrow&
p_a^{\mu_1}M^{(2)}_{\mu_1\nu_2}
d_{\mu_2}^{\nu_2}(q_1)p_b^{\mu_2}\nonumber \\
p_a^{\mu_1}M^{(3)}_{\mu_1\mu_2}p_b^{\mu_2}
&\rightarrow&
p_a^{\mu_1}d_{\mu_1}^{\nu_1}(q_1)M^{(3)}_{\nu_1\mu_2}
p_b^{\mu_2},\label{d-subst}\\
p_a^{\mu_1}M^{(4)}_{\mu_1\mu_2}p_b^{\mu_2}
&\rightarrow&
p_a^{\mu_1}M^{(4)}_{\mu_1\mu_2}p_b^{\mu_2},\nonumber
\end{eqnarray}
where $M^{(i)}$ have to be calculated in the new gauge.
Now one can present $M^{(2)}$, $M^{(3)}$ and $M^{(4)}$ as follows
\ba
\tilde{M}^{(2)}_{\mu_1\mu_2}&=&
-\frac{q_{1\perp, \mu_1}}{x_{2}}\frac{p_a^{\nu_1}}{S}
M^{(2)}_{\nu_1\mu_2}\nonumber\\
\tilde{M}^{(3)}_{\mu_1\mu_2}&=&
-\frac{q_{2\perp, \mu_2}}{x_{1}}
M^{(3)}_{\mu_1\nu_2}\frac{p_b^{\nu_2}}{S}\nonumber\\
\tilde{M}^{(4)}_{\mu_1\mu_2}&=&
-\frac{q_{1\perp, \mu_1}}{x_{2}}\frac{q_{2\perp, \mu_2}}{x_{1}}
\frac{p_a^{\nu_1}}{S}M^{(4)}_{\nu_1\nu_2}\frac{p_b^{\nu_2}}{S}\nonumber
\ea
Using now (\ref{projection}) we obtain
$$
p_a^{\mu_1}M_{\mu_1\mu_2}p_b^{\mu_2}\rightarrow
\frac{q_{1\perp}^{\mu_1}}{x_{1}}\tilde{M}_{\mu_1\mu_2}
\frac{q_{2\perp}^{\mu_2}}{x_{2}},
$$
where
$$
\tilde{M}=M^{(1)}+\tilde{M}^{(2)}+\tilde{M}^{(3)}+\tilde{M}^{(4)}.
$$

The amplitude $A_{2\rightarrow n+2}$ projected onto the physical states
of incoming and outgoing particles is, of course, gauge invariant.
While gauge transformations do not change $\Gamma_{a,b}$
(when transforming from the covariant to the axial gauge)
the $p_a^{\mu_1}M_{\mu_1\mu_2}p_b^{\mu_2}$ projected onto the physical
polarizations of outgoing particles also remains the same
 (see (\ref{d-subst})). This proves one can rewrite $M_{\mu_1\mu_2}$
in the form where the t-channel gluons having  momenta $q_{1\perp}$ and
$q_{2\perp}$ are having transverse polarizations in the original covariant gauge
when the amplitude is projected onto the physical subspace.

\appex{Cross sections of pair production in high energy factorization}

Let us introduce the following notations
$$
s=2(k_1k_2\ch(\Delta y)-k_{1\perp}k_{2\perp});
$$
$$
t=-(q_{1\perp}-k_{1\perp})^2-k_1k_2e^{\Delta y},\;
u=-(q_{1\perp}-k_{2\perp})^2-k_1k_2e^{-\Delta y};
$$
$$
\Sigma=x_1x_2S=k_1^2+k_2^2+2k_1k_2\ch(\Delta y),
$$
where  $k_1=\sqrt{k_{1\perp}^2}$, $k_2=\sqrt{k_{2\perp}^2}$
and $k_{1\perp}k_{2\perp}$ is a dot product with 2d Euclidean metric.

The combined contribuion from gluons and quarks (fermions) to $gg$
scattering has the form (see (\ref{sigma2}))
\be\label{Atotal}
{\cal A}={\cal A}_{gluons}+\frac{n_f}{4N^3_c}{\cal A}_{fermions}
\ee

\subsection{$gg\rightarrow gg$}
$$
{\cal A}_{gluons}={\cal A}_1+{\cal A}_2
$$

\begin{eqnarray}
{\cal A}_1=
&q_1^2q_2^2&
    \left\{ -\frac{1}{tu}+\frac{1}{4tu}\frac{q_1^2q_2^2}{k_1^2k_2^2}-
            \frac{e^{\Delta y}}{4tk_1k_2}-\frac{e^{-\Delta y}}{4uk_1k_2}+
            \frac{1}{4k_1^2k_2^2}+
    \right.\nonumber\\
&&         \frac{1}{\Sigma}
             \left[-\frac{2}{s}
                \left(1+k_1k_2(\frac{1}{t}-\frac{1}{u})\sh(\Delta y)
                \right)+
                \frac{1}{2k_1k_2}(1+\frac{\Sigma}{s})\ch(\Delta y)-
             \right.\nonumber\\
&&
               -\frac{q_1^2}{4s}
                      [(1+\frac{k_2}{k_1}e^{-\Delta y})\frac{1}{t}+
                       (1+\frac{k_1}{k_2}e^{\Delta y})\frac{1}{u}]\nonumber\\
&&           \left.
    \left.
               -\frac{q_2^2}{4s}
                      [(1+\frac{k_1}{k_2}e^{-\Delta y})\frac{1}{t}+
                       (1+\frac{k_2}{k_1}e^{\Delta y})\frac{1}{u}]
             \right]
    \right\}
\label{gggg}
\end{eqnarray}
\begin{eqnarray}
{\cal A}_2=\frac{1}{2}\left\{\left(
\frac{(k_{1\perp}-q_{1\perp})^2
(k_{2\perp}-q_{1\perp})^2-k_1^2k_2^2}{tu}\right)^2-
\right.\nonumber\\
-\frac{1}{4}
\left.\left( \frac{(k_{2\perp}-q_{1\perp})^2-k_1k_2e^{-\Delta y}}
                  {(k_{2\perp}-q_{1\perp})^2+k_1k_2e^{-\Delta y}}
             -\frac{E}{s}
      \right)
      \left( \frac{(k_{1\perp}-q_{1\perp})^2-k_1k_2e^{\Delta y}}
                  {(k_{1\perp}-q_{1\perp})^2+k_1k_2e^{\Delta y}}
             +\frac{E}{s}
      \right)
\right\},\nonumber
\end{eqnarray}

$$
E=(q_{1\perp}-q_{2\perp})(k_{1\perp}-k_{2\perp})-\frac{1}{\Sigma}
(q_1^2-q_2^2)(k_1^2-k_2^2)+2k_1k_2\sh(\Delta y)
\left(1-\frac{q_1^2+q_2^2}{\Sigma}\right).
$$

\subsection{$gg\rightarrow q\bar{q}$}
$$
{\cal A}_{fermions}=N_c^2{\cal A}_{1f}+{\cal A}_{2f}
$$

\begin{eqnarray}
{\cal A}_{1f}&=&\left\{ 2\frac{q_{1}^{2}q_{2}^{2}}{s\Sigma }
\left(1+k_{1}k_{2}\sh (\Delta y)(\frac{1}{t}-\frac{1}{u})\right)
-\left( \frac{(k_{1\perp}-q_{1\perp})^{2}(k_{2\perp}-q_{1\perp})^{2}
-k_{1}^{2}k_{2}^{2}}{tu}\right)^{2}+\right.   \nonumber \\
&&\left. \frac{1}{2}
      \left( \frac{(k_{2\perp}-q_{1\perp})^2-k_1k_2e^{-\Delta y}}
                  {(k_{2\perp}-q_{1\perp})^2+k_1k_2e^{-\Delta y}}
             -\frac{E}{s}
      \right)
      \left( \frac{(k_{1\perp}-q_{1\perp})^2-k_1k_2e^{\Delta y}}
                  {(k_{1\perp}-q_{1\perp})^2+k_1k_2e^{\Delta y}}
             +\frac{E}{s}
      \right)
\right\}\label{ggqq}
\end{eqnarray}
and
$$
{\cal A}_{2f}=\left\{ \left( \frac{(k_{1\perp}-q_{1\perp})^{2}
(k_{2\perp}-q_{1\perp})^{2}-k_{1}^{2}k_{2}^{2}}{tu}\right)^{2}
-\frac{q_{1}^{2}q_{2}^{2}}{tu}\right\}
$$
where $E$ is the same as for gluons.

In multy-Regge kinematics only term leading in $\Delta y\rightarrow\infty$
survives and
$$
{\cal A}_{MRK}={\cal A}_{MRK, gluons}=\frac{q_1^2q_2^2}{k_1^2k_2^2}
$$

\newpage

\begin{center}
\large\bf Figure captions
\end{center}

The cross sections depicted in figures are calculated for
unintegrated structure functions CTEQ5M \cite{CTEQ} and AKMS
\cite{AKMS} with constant $\alpha_s=0.2$. Gluon and quark
contributions (with $n_f=5$) are added up.

\begin{figure}
\begin{center}
\hspace*{-0.5cm}
\epsfig{file=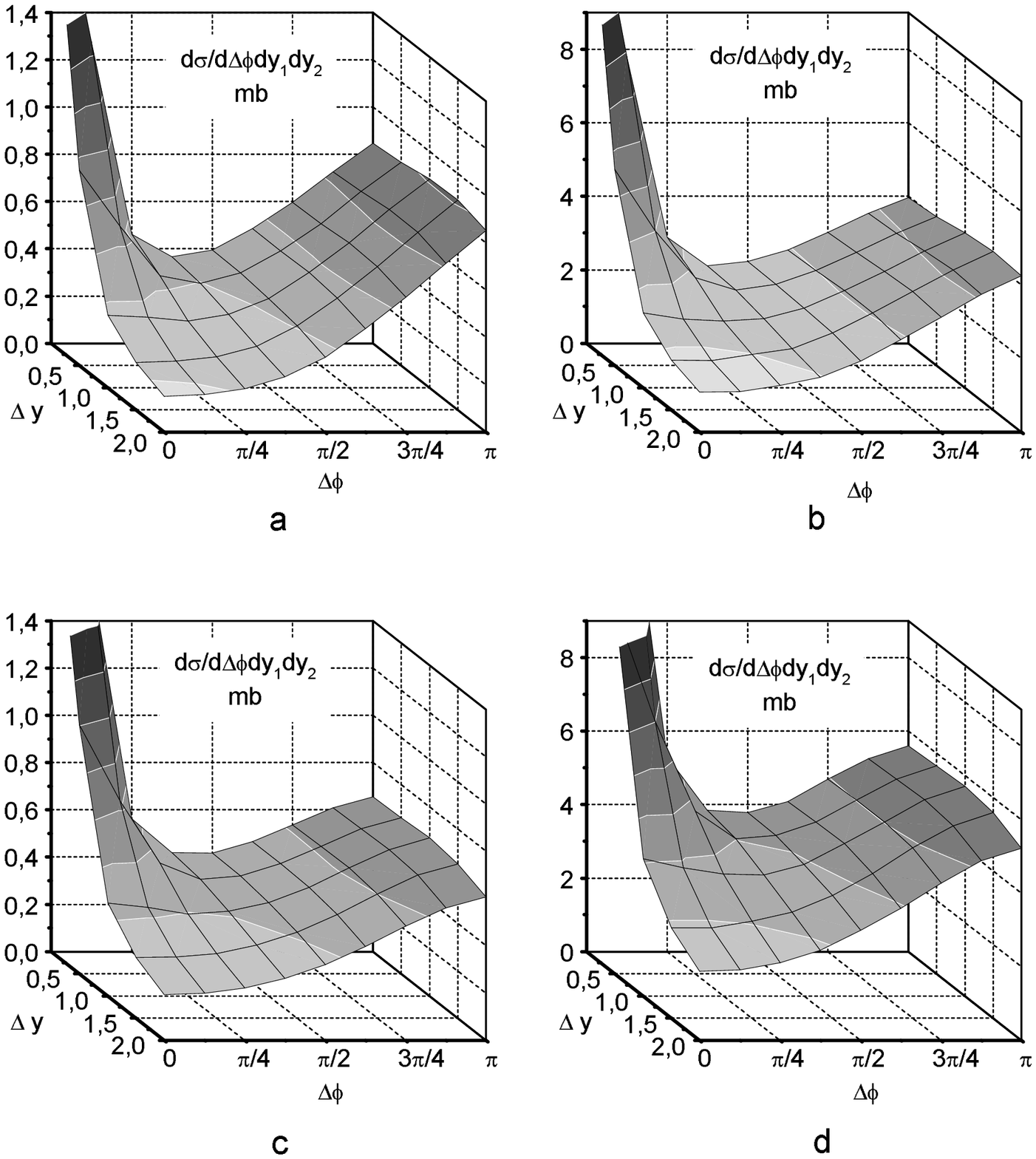,width=16cm}
\vspace*{0cm}
\figcaption\label{ff3d}
\end{center}
\end{figure}

\begin{figure}

\begin{center}
\hspace*{-1cm}
\epsfig{file=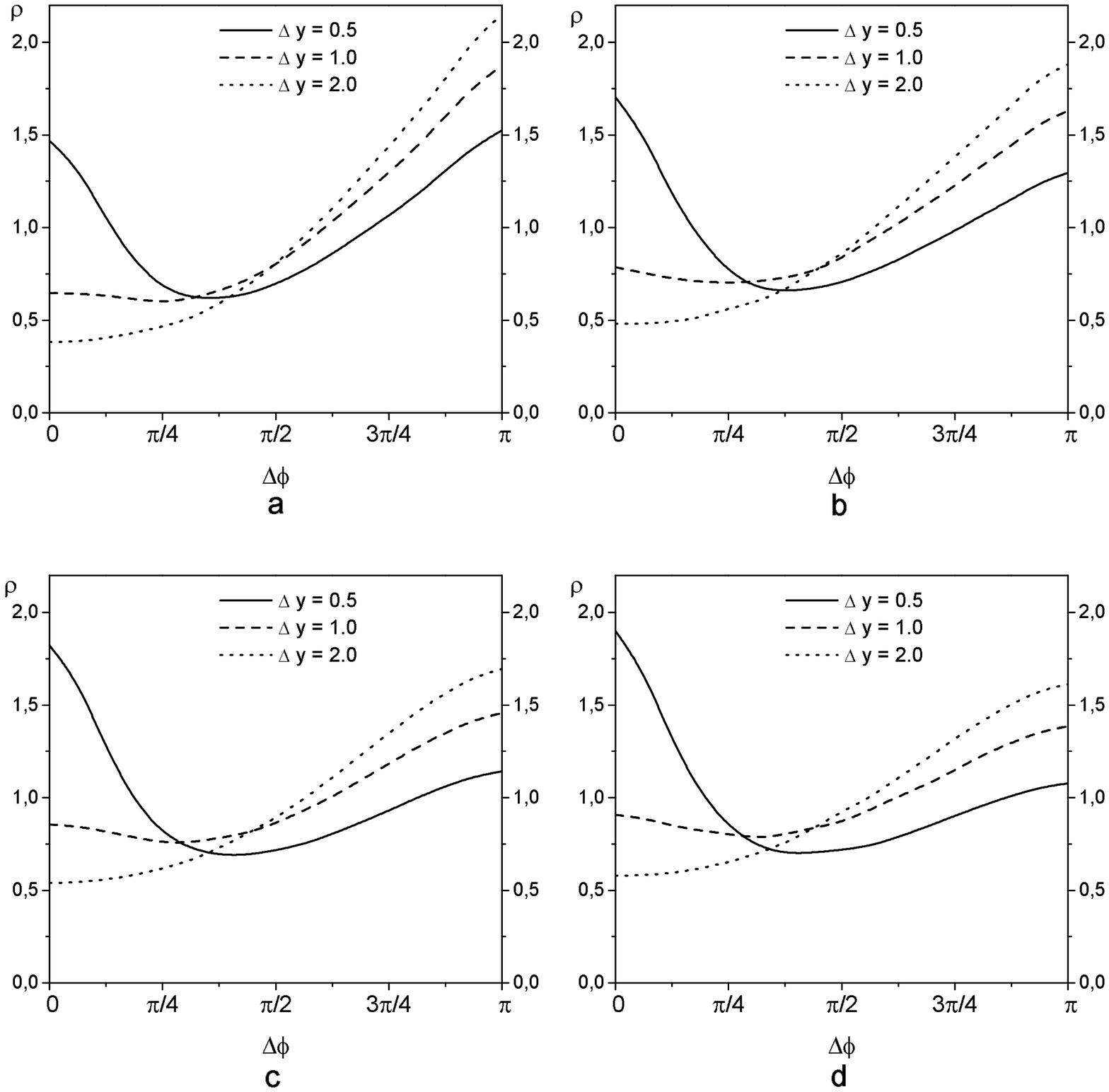,width=17cm}
\vspace*{0.5cm}
\figcaption\label{rhoy}
\end{center}
\end{figure}

\begin{figure}
\vspace*{-1.5cm}
\begin{center}
\hspace*{-0.5cm}
\epsfig{file=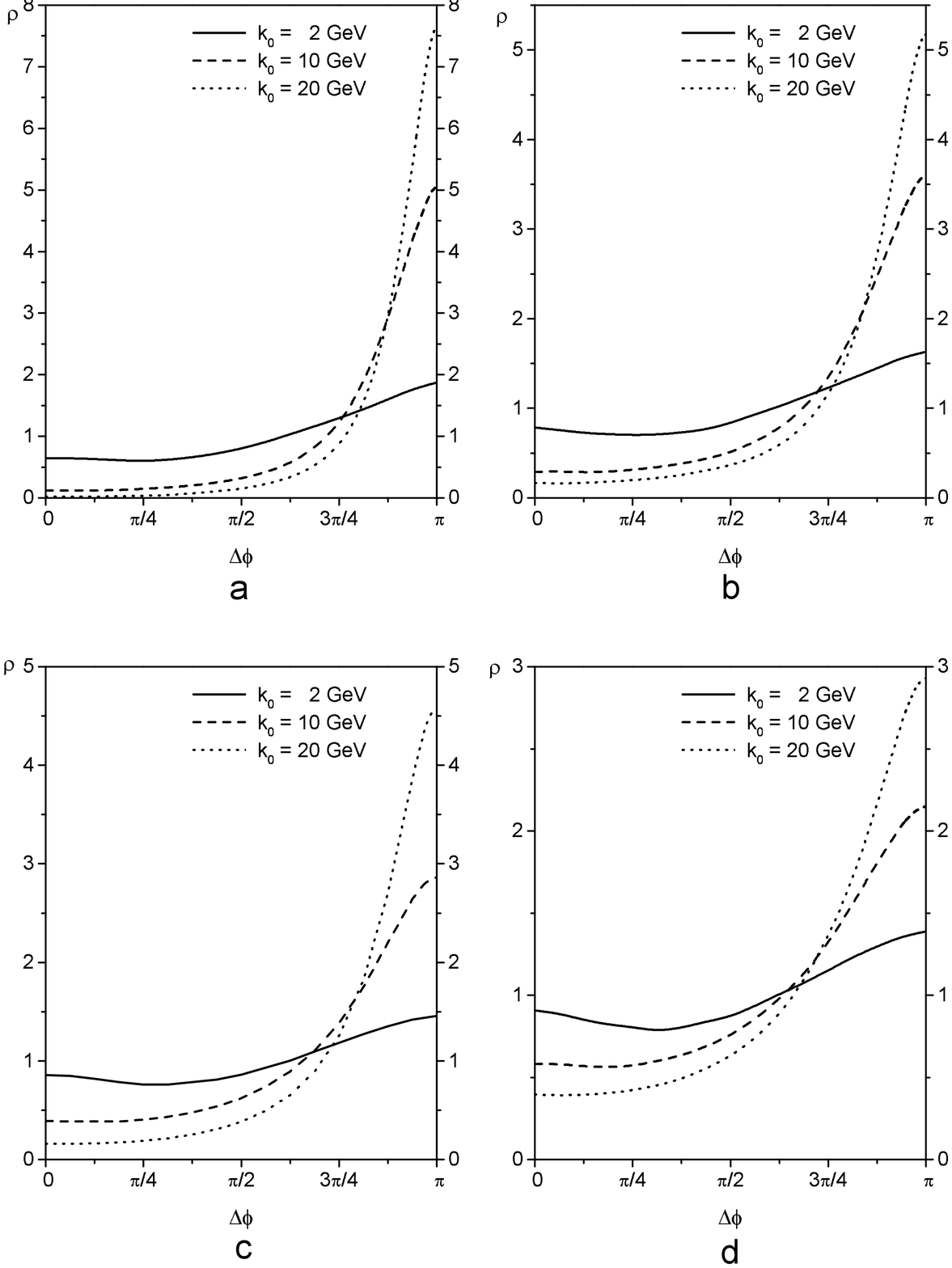,width=165mm}
\vspace*{0cm}
\figcaption\label{rhok}
\end{center}
\end{figure}

\begin{figure}

\begin{center}
\hspace*{-10mm}
\epsfig{file=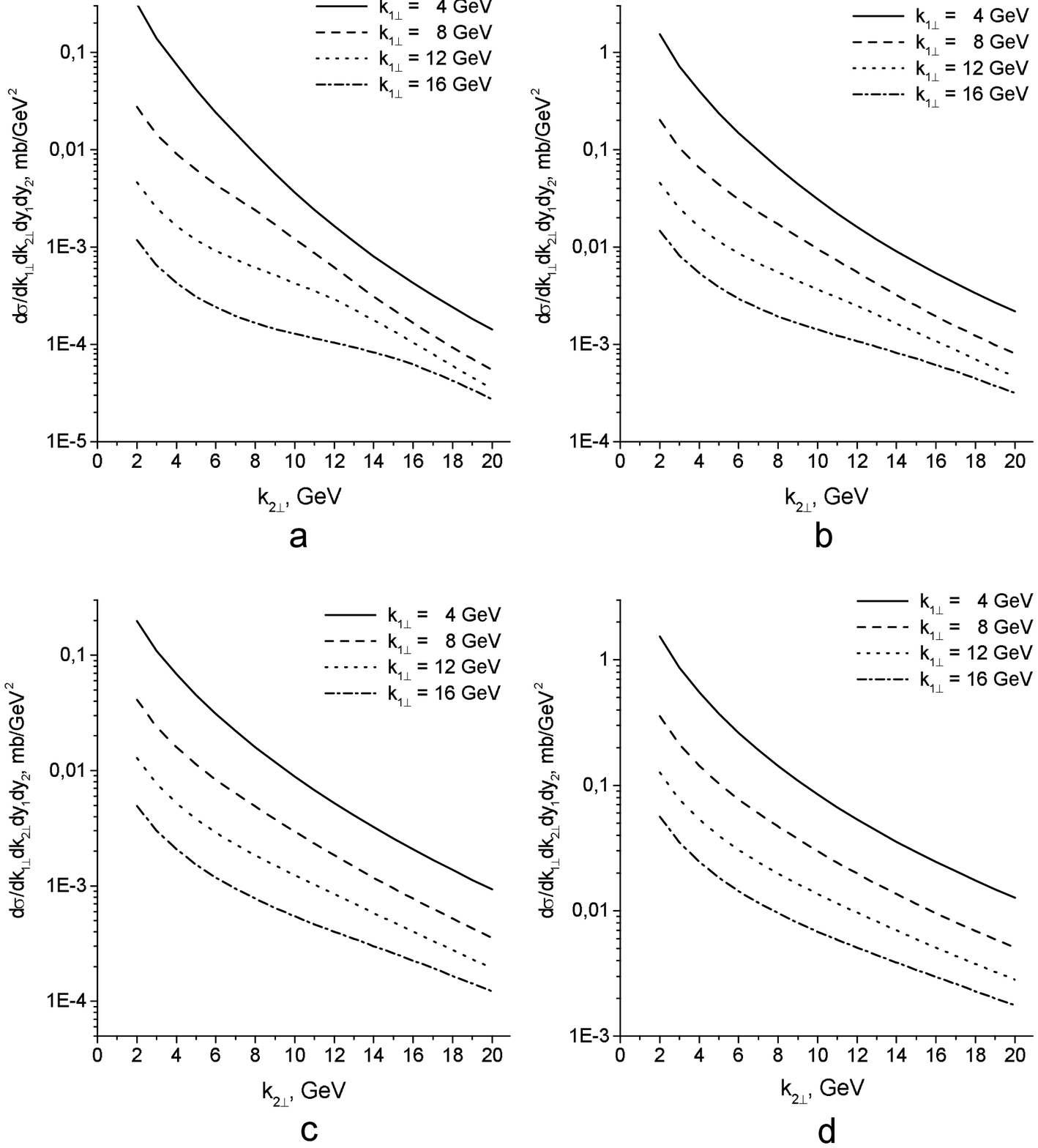,width=17cm}
\vspace*{0cm}
\figcaption\label{mm}
\end{center}
\end{figure}

\begin{itemize}

\figitem{ff3d} Cross section Eq.~(\ref{angle})
calculated for $k_0=2\,{\rm GeV}$, $y_0=0$ and $\Delta y\in [0,2]$\\
{\bf a} with CTEQ5M structure functions, $\sqrt{S}=1.8\,{\rm TeV}$\\
{\bf b} with CTEQ5M structure functions, $\sqrt{S}=14\,{\rm TeV}$\\
{\bf c} with AKMS structure functions, $\sqrt{S}=1.8\,{\rm TeV}$\\
{\bf d} with AKMS structure functions, $\sqrt{S}=14\,{\rm TeV}$

\figitem{rhoy} Angular asymmetry $\rho(\Delta\phi)$ for
$k_0=2\,{\rm GeV}$, $y_0=0$ and $\Delta y=0.5, 1, 2$.
Specific parameters for {\bf a, b, c, d} same as above.

\figitem{rhok} Angular asymmetry $\rho(\Delta\phi)$ for
$y_{1,2}=\pm 0.5$ and $k_0=4, 8, 12, 16\, {\rm GeV}$.
Specific parameters for {\bf a, b, c, d} are the same as earlier.

\figitem{mm} Momentum asymmetry Eq.~(\ref{momentum}) for $y_{1,2}=\pm 0.5$.
Specific parameters for {\bf a, b, c, d} same as above.

\end{itemize}

\newpage
\textwidth=17cm
\topmargin=-10mm

\newpage

\newpage

\end{document}